\documentclass[aip,jcp,reprint,groupedaddress,twocolumn]{revtex4}
\usepackage{graphicx}
\usepackage{epsfig}
\usepackage{amsmath}
\usepackage{amssymb}
\usepackage{hhline}

\begin{document}

\title{Stability  analysis of multiple  nonequilibrium fixed points in self-consistent  electron transport calculations }

\author{Alan A. Dzhioev}
\altaffiliation{Permanent address: Bogoliubov Laboratory of Theoretical Physics, Joint Institute for Nuclear Research,  RU-141980 Dubna, Russia }
\author{D. S. Kosov}
\email{dkosov@ulb.ac.be}
\affiliation{Department of Physics,
Universit\'e Libre de Bruxelles, Campus Plaine, CP 231, Blvd du Triomphe, B-1050 Brussels, Belgium }

\pacs{05.30.-d, 05.60.Gg, 72.10.Bg}

\begin{abstract}
We present a method to perform stability analysis  of nonequilibrium fixed points appearing in self-consistent  electron transport calculations. The nonequilibrium fixed points are given by the self-consistent solution of stationary, nonlinear kinetic equation for single-particle density matrix. We obtain the stability matrix by linearizing the kinetic equation around the fixed points and
analyze the real part of its spectrum to assess the asymptotic time behavior of the fixed points.
We derive expressions for the stability matrices within Hartree-Fock and linear response adiabatic time-dependent density functional theory.
The stability analysis of multiple fixed points is performed within the nonequilibrium Hartree-Fock approximation for the electron transport through a molecule with a spin-degenerate single level with local Coulomb interaction.
\end{abstract}

\maketitle

\section{Introduction}

The existence of nonunique steady state for nonequilibrium systems of correlated quantum or classical particles is
an interesting and open  fundamental problem.\cite{dhar-rev,leeuwen10}
Multiple steady states in nanojunctions  lead to bistabilities and  hysteresis loops in the current-voltage characteristics
and this  currently  is a much debated theoretical issue.\cite{PhysRevB.67.075301,galperin05,negre}
Density  functional theory (DFT) and Hartree-Fock based nonequilibrium Green's functions (NEGF) electron transport calculations, which are widely used nowadays, solve nonlinear system of equations for nonequilibrium electron density via
self-consistent iterations.\cite{Taylor01,Brandbyge02,Xue02,Ke04}
Such kind of nonlinear problems may have multiple solutions.\cite{sanchez,negre} In equilibrium case  the correct physical solution corresponds to a minimum of the ground state energy.
The situation is less clear in  nonequilibrium  where the system is open and the minimum energy arguments are not applicable anymore.
In this paper, we discuss the appearance of multiple fixed points
and  present a method to eliminate unphysical solutions of steady-state nonequilibrium self-consistent electron transport problem.

Let us consider a finite quantum system  (e.g., a molecule) connected to two macroscopic particle reservoirs or thermal baths (e.g., metal electrodes).
By projecting out bath degrees of freedom we obtain the kinetic equation for the reduced density matrix $\rho(t)$  of the embedded system
 \begin{equation}
i \frac{d}{dt} \rho(t) = {\cal L} \rho(t),
   \label{liouville}
 \end{equation}
 where $  {\cal L} $ is a non-Hermitian Liouvillian.  We assume that the dynamics is markovian and the system is autonomous, i.e.  $  {\cal L} $ does not depend explicitly on time.
 In this paper we will focus on nonequilibrium steady state.
    Like an equilibrium represents stationary state of a closed system,  a nonequilibrium steady state is the {\it stable}, time-invariant state of an open system. If the dynamics  generated by ${\cal L} $ is linear, then a nonequilibrium steady state  can be unambiguously
defined as a state when the left side of the equation (\ref{liouville}) becomes zero.  However,
often times for practical calculations we involve the mean-field approximation (Hartree-Fock or DFT) and  the Liouvillian becomes a functional of the reduced density matrix
 \begin{equation}
  {\cal L} = {\cal L} [\rho],
 \end{equation}
Therefore the kinetic equation (\ref{liouville}) becomes nonlinear and the issue of stability of the solution becomes pivotal.

Let us give a few formal definitions from the theory of dynamical dissipative systems,\cite{mikhailov,Strogatz}
which are relevant to electron transport problem.
The density matrices $\overline{\rho}$ at which
 \begin{equation}
 {\cal L} [\overline{\rho}] \overline{\rho} =0,
 \label{steady}
 \end{equation}
are called the {\it fixed points} of the system.
 Since Eq.(\ref{steady}) is nonlinear, it generally has multiple solutions, which may or may not be
steady state (i.e. stable fixed point).
To understand whether or not the fixed point is stable we need to perform stability analysis commonly used  for dynamical nonlinear systems. We expand the density around the fixed point
\begin{equation}
\rho(t) = \overline{\rho} + \delta \rho(t)
\end{equation}
and linearize the Liouville equation
 \begin{equation}
\frac{d}{dt} {\delta \rho}(t) =A  \delta \rho(t),
\label{linear}
 \end{equation}
 where
 \begin{equation}
 A=   -i \left[ {\cal L}[\overline{\rho}]  + \left. \frac{\delta {\cal L} }{\delta \rho} \right|_{\rho=\overline{\rho}}  \overline{\rho} \right]
 \end{equation}
 is the stability matrix.
 If $A$ is a so-called Hurwitz matrix, i.e., if  all eigenvalues $\lambda_i$ of $A$ satisfies the conditions $\text{Re}(\lambda_i) <0$, then the fixed point $\overline{\rho}$ is asymptotically stable and it is the true steady state of the system. If at least one eigenvalue has positive real part, the solution is unstable and can not be a steady state, since even an infinitesimally small variation of the fixed point density matrix drives the solution away from the fixed point exponentially in time. In our paper we demonstrate that there are several fixed points and multiple steady states in some rather typical cases of electronic transport calculations.

The rest of the paper is organized as follows. In Section II,  we discuss nonequilibrium fixed points, linearization and stability matrix
for self-consistent electron transport problem and obtain the general expression for the stability matrix.
In section III, we apply the method to out of equilibrium Anderson model and demonstrates that it has multiple (stable and unstable) fixed points.
Conclusions are given in Section IV. In Appendix A, we derive the kinetic equation for
the reduced density matrix of the embedded system.  The adiabatic  time-dependent Kohn-Sham DFT expressions  for stationary Fock matrix and its fluctuating part are  given in ppendix B.
We use natural units throughout the paper: $\hbar= k_B = |e| = 1$, where $-|e|$ is the electron charge.

\section{Nonequilibrium fixed points, linearization and stability matrix}

Let us consider a  molecule connected to two electrodes.
We partition the system into five parts: the molecule itself, the left/right macroscopically large leads (environment),  and the left/right finite buffer zones between the
molecule and the environment.
The Hamiltonian is written in the following form:
\begin{equation}
{\cal H} = H_{M}+ H_E + H_B + H_{MB} + H_{EB}.
\label{h}
 \end{equation}
The environment and the buffer zones are described by the noninteracting Hamiltonians
\begin{equation}
{  H}_E = \sum_{\sigma,  k\in L,R  }  \varepsilon_{k } a^{\dagger}_{ k \sigma } a_{ k  \sigma },
\end{equation}
\begin{equation}
{  H}_B = \sum_{\sigma, b \in L,R}  \varepsilon_{ b } a^{\dagger}_{b \sigma }  a_{b \sigma }.
\end{equation}
Here $\varepsilon_{k}$ denote the continuum single-particle spectra of the  left ($k\in L$) and  right ($k\in R$)   lead states,
 $a^{\dagger}_{k \sigma }$ ($a_{k \sigma }$) create (annihilate) electron with spin $\sigma =\uparrow, \downarrow$ in the lead state
$ {k }$.
The buffer zones have discrete energy spectrum $\varepsilon_{ b }$ with corresponding creation and annihilation operators
$a^\dag_{b \sigma }$ and $a_{b \sigma}$.

The  molecular Hamiltonian is taken in the most general form and contains electronic kinetic energy, electron-ion interaction, and Coulomb interaction between electrons:
  \begin{equation}
   H_M = \sum_{ ij \sigma } T_{ij}  a^\dag_{i \sigma} a_{j \sigma} +\frac{1}{2} \sum_{ijmn} \sum_{\sigma \sigma'}(ij|mn)  a^\dag_{i \sigma}  a^\dag_{m \sigma'} a_{n \sigma'}  a_{j \sigma},
\end{equation}
where
\begin{equation}
T_{n m}  = \int d{\bf r} \phi_{n} ({\bf r}) \Bigl( -\frac{1}{2} \nabla^2 + V_{\text{e-i}}({\bf r})\Bigr)  \phi_{m} ({\bf r})
\end{equation}
and
\begin{equation}
(i j | m n)   = \int d{\bf r} d{\bf r}'   \phi_{i} ({\bf r})  \phi_{j} ({\bf r}) \frac{1}{|{\bf r} -{\bf r}'|} \phi_{m} ({\bf r}')  \phi_{n} ({\bf r}')
\end{equation}
are matrix elements computed in some orthogonal real basis $\langle \phi_i | \phi_j \rangle = \delta_{ij} $.
Here $a^\dag_{i \sigma}$ and $a_{i \sigma}$ are creation and annihilation operators for electron with spin $\sigma$ in molecular
state $|\phi_i\rangle$.
Hereinafter, the index $b$  refers to   discrete single-particle states in either  left  or right buffer zones,
whereas the indices $i,j,m,n$ represent states in the molecular space.

The buffer-environment and molecule-buffer coupling  have the standard tunneling form:
\begin{equation}
{  H}_{EB} = \sum_{\sigma, b k \in L }  ( v_{b k } a^{\dagger}_{b  \sigma  }  a_{k \sigma  } + h.c.) +
\sum_{\sigma, b k \in R }  ( v_{b k } a^{\dagger}_{b  \sigma  }  a_{k \sigma  } + h.c.),
\label{v}
\end{equation}
\begin{equation}
{H}_{MB} = \sum_{ i \sigma,  b \in L,R  }  ( t_{ i b \sigma} a^{\dagger}_{b \sigma } a_{i \sigma } + h.c. ).
\label{t}
\end{equation}

The Liouville equation for the total density matrix $\chi(t)$ is:
\begin{equation}
i \dot{{\chi}}(t) = [{\cal H}, \chi(t)].
\end{equation}
If we project out  the environment degrees of freedom
and make the standard assumptions (Born-Markov and rotating wave approximations\cite{Petruccione} -- the details of the derivation are shown in Appendix A)
we get the following kinetic equation
\begin{eqnarray}
i \dot \rho(t) =  [ H,  \rho(t) ]+ \hat \Pi \rho(t)
 \label{lindblad}
\end{eqnarray}
for the reduced density matrix for the embedded molecule (i.e. for the
molecule and the buffer zones)
\begin{equation}
\rho(t) = \text{Tr}_E  \chi(t).
\end{equation}
Here the Hamiltonian $H$ includes the Lamb shift of the single-particle levels of  the buffer zones
\begin{equation}
H= H_M + H_{MB} + \sum_{\sigma, b } (\varepsilon_{b} + \Delta_{b} )  a^{\dagger}_{b \sigma} a_{b \sigma},
\end{equation}
and the non-Hermitian dissipator is given by standard Lindblad form
\begin{align}\label{non_herm}
 \hat{\Pi}\rho(t) = \sum_{\sigma, b} \sum_{\mu=1,2} \bigl(2 L_{ b\sigma   \mu}\rho(t) L^\dag_{ b\sigma  \mu} - \{L^\dag_{b \sigma \mu} L_{ b \sigma  \mu},\rho(t) \}\bigr)
\end{align}
with the following  Lindblad operators
\begin{align}
  L_{b\sigma 1} = \sqrt{\gamma_{b }(1-f_b)}a_{b \sigma},~~L_{b \sigma  2} = \sqrt{\gamma_{b}  f_b}a^\dag_{ b  \sigma}.
\end{align}
Here $\Delta_{b} $ and $\gamma_{b } $ are real and imaginary parts of the standard environment self energy
$\sum_k |v_{bk }|^2/(\varepsilon_{b }-\varepsilon_{k } +i0^+)$ and $ f_{b\in L/R} =[1+ e^{\beta_{L/R}(\varepsilon_{b} - \mu_{L/R}) }]^{-1}$.

We emphasize  that our kinetic equation (\ref{lindblad}) does not employ the second order perturbative treatment of molecule-electrode coupling (\ref{t}), but rather it is the second-order in terms of the coupling between the buffer zone and the environment (\ref{v}).
We have recently demonstrated that
for the steady
state electron transport calculations the kinetic equation (\ref{lindblad})
can be made as accurate and exact as necessary in practical calculations by
increasing the density of single-particle buffer states $b$ included into
the Hamiltonian.\cite{dzhioev11a,dzhioev11b}
The similar idea of the buffer zone between the molecule and the environment  has been recently proposed in dynamical simulations of inelastic electron transport.\cite{todorov07}

Let us  now consider  time-evolution of the expectation value of an arbitrary operator $O$:
\begin{equation}
\langle  O  \rangle_t = \text{Tr} [ \rho(t) O ].
\end{equation}
Using Eq.~\eqref{lindblad} we obtain
\begin{align}
&\frac{d}{dt} {\langle  O  \rangle_t}  = -i \langle [ O, H ] \rangle_t
  \notag  \\
  +
&  \sum_{ \sigma b} \sum_{\mu=1,2}\bigl( 2 \langle L^{\dag}_{ b \sigma \mu} O L_{ b \sigma \mu} \rangle_t
  - \langle \{L^\dag_{b \sigma \mu} L_{ b \sigma \mu}, O \} \rangle_t\bigr).
 \label{lindblad-op}
\end{align}
We apply the time-dependent Hartree-Fock approximation to the two-particle interaction in the molecular Hamiltonian
\begin{equation}
H^{(0)}_M = \sum_{\sigma ij} F^\sigma_{ij}(t) a^{\dag}_{i \sigma} a_{j \sigma}.
\label{hf-hamiltonian}
\end{equation}
Here $ F^{\sigma}_{nm} (t)$ is the time-dependent Fock matrix
\begin{equation}
F^{\sigma}_{nm} (t) = T_{nm} +  \sum_{ij} \left[ (n m | ij) P_{  ij}(t) - (n i | m j) P^{\sigma}_{ ij}(t)\right],
\label{fock-matrix}
\end{equation}
which depends on the single-particle density matrices
\begin{equation}
P^{\sigma}_{ ij}(t) =\langle  a^{\dag}_{j \sigma} a_{i \sigma} \rangle_t \; , \qquad
P_{ij}(t) = \sum_{\sigma} P^{\sigma}_{ i j}(t).
\end{equation}
The corresponding expression for the Fock matrix in  adiabatic time-dependent Kohn-Sham DFT is given in Appendix B.
The Lindblad equation (\ref{lindblad}) becomes the time-dependent Hartree-Fock equation for the nonequilibrium  open quantum system, when the full many-body molecular Hamiltonian  is approximated by $H^{(0)}_M$ (\ref{hf-hamiltonian}).
We  introduce two additional single-particle density matrices:
\begin{equation}
P^{\sigma}_{ bj}(t) = \langle a^{\dag}_{j \sigma} a_{b \sigma} \rangle_t \; , \qquad  P^{\sigma}_{ bb'}(t) = \langle  a^{\dag}_{b'\sigma} a_{b \sigma} \rangle_t
\end{equation}
and $P^{\sigma}_{ jb}(t) = (P^{\sigma}_{ bj}(t))^*$.
By means of  (\ref{lindblad-op}) we get the close set of time evolution equations for single-particle density matrices
\begin{widetext}
\begin{eqnarray}
i \frac{d}{dt}  P^{\sigma}_{ ij}(t)  &=& \sum_n \bigl[ F^{\sigma}_{in}(t) P^\sigma_{nj}(t) - F^{\sigma}_{nj}(t) P^\sigma_{in}(t)\bigr] + \sum_b\bigr[t_{ib\sigma} P^\sigma_{bj}(t)- t^*_{jb\sigma}P^\sigma_{ib}(t)\bigl],
       \notag    \\
i\frac{d}{dt}   P^{\sigma}_{ bi}(t) &=&  E_b P^{\sigma}_{ bi}(t)  - \sum_j  F^{\sigma}_{ji}(t) P^{\sigma}_{ bj}(t) - \sum_{b'} t^*_{ib'\sigma}P^{\sigma}_{ b'b}(t) + \sum_j t^*_{jb\sigma} P^\sigma_{ji}(t),
      \notag \\
i \frac{d}{dt}    P^{\sigma}_{ bb'}(t) &=& (E_b - E^*_{b'}) P^{\sigma}_{ bb'}(t) - \sum_{i}\bigr[t_{ib'\sigma} P^\sigma_{bi}(t) - t^*_{ib\sigma} P^\sigma_{ib'}(t)\bigl] + 2i\delta_{bb'} f_b\gamma_b
\label{tdhf}
\end{eqnarray}
\end{widetext}
with $E_b = \varepsilon_b - i\gamma_b$. Here we include the Lamb shift $\Delta_b$ into single-particle energy~$\varepsilon_b$.
These equations of motion are nonlinear because the Fock matrix $F^\sigma_{ij}(t)$ depends on the density matrix $P^\sigma_{ij}(t)$.

Setting the left  side of equations \eqref{tdhf} to zero, we obtain the stationary  Hartree-Fock equations for  nonequilibrium fixed points.
These fixed points correspond to stationary single-particle densities,
$\overline{P}^{\sigma}_{ ij}$, $ \overline{P}^{\sigma}_{ bi}$,$ \overline{P}^{\sigma}_{ ib}$, and $ \overline{P}^{\sigma}_{ bb'}$,
which may or may not be steady state densities. To determine  if these fixed points are asymptotically stable, we  linearize Eq.~\eqref{tdhf} around each fixed point.
Substituting (here indices $\alpha, \beta$ run over molecular and buffer single particle states)
\begin{equation}
 P^{\sigma}_{ \alpha \beta}(t)  = \overline{P}^{\sigma}_{ \alpha \beta } + \delta P^{\sigma}_{ \alpha \beta}(t)
\label{delta}
\end{equation}
into Eq.~\eqref{tdhf} and retaining only the terms linear in  $\delta P^{\sigma}_{ \alpha \beta}(t)$  we get
\begin{widetext}
\begin{eqnarray}
i \frac{d}{dt}   {\delta P}^{\sigma}_{ ij}(t)  &=& \sum_n \bigl[ \overline{F}^{\sigma}_{in} \delta P^\sigma_{nj}(t) - \overline{F}^{\sigma}_{nj} \delta P^\sigma_{in}(t)
+ \delta F^{\sigma}_{in}(t) \overline{P}^\sigma_{nj} - \delta F^{\sigma}_{nj}(t) \overline{P}^\sigma_{in}
\bigr] + \sum_b\bigr[t_{ib\sigma} \delta P^\sigma_{bj}(t) - t^*_{jb\sigma}\delta P^\sigma_{ib}(t)\bigl],
      \notag    \\
i\frac{d}{dt}  {\delta P}^{\sigma}_{ bi}(t) &=&  E_b \delta P^{\sigma}_{ bi}(t)  - \sum_j \bigl[ \overline{F}_{ji} \delta P^{\sigma}_{ bj}(t)+ \delta F^{\sigma}_{ji}(t) \overline{P}^{\sigma}_{ bj} \bigr]- \sum_{b'} t^*_{ib'\sigma} \delta P^{\sigma}_{ b'b}(t) + \sum_j t^*_{jb\sigma} \delta P^\sigma_{ji}(t),
  \notag \\
i \frac{d}{dt}   {\delta P}^{\sigma}_{ bb'}(t) &=& (E_b - E^*_{b'}) \delta P^{\sigma}_{ bb'}(t) - \sum_{i}\bigr[t_{ib'\sigma} \delta P^\sigma_{bi}(t) - t^*_{ib\sigma} \delta P^\sigma_{ib'}(t)\bigl].
\label{tdhf-lin}
\end{eqnarray}
\end{widetext}
Here the Fock matrix at the fixed point  is
\begin{equation}
 \overline{F}^{\sigma}_{nm} = T_{nm} +  \sum_{ij} \left[ (n m | ij) \overline{P}_{  ij}- (n i | m j) \overline{P}^{\sigma}_{ ij}\right].
 \label{fst}
\end{equation}
and its time-dependent fluctuation around this fixed point is
\begin{equation}
\delta F^{\sigma}_{nm}(t) =  \sum_{ij} \left[ (n m | ij) \delta P_{  ij}(t) - (n i | m j) \delta P^{\sigma}_{ ij}(t)\right].
\label{deltaf}
\end{equation}
The adiabatic  time-dependent Kohn-Sham DFT expressions  for stationary Fock matrix and its fluctuating part are  given in appendix B.

The system of equations (\ref{tdhf-lin}) can be rewritten as the set of linear differential equation
\begin{equation}
\frac{d}{dt}  {\delta P}^{\sigma}_{\alpha \beta}(t)=\sum_{\alpha' \beta'} \sum_{\sigma'} A^{\sigma \sigma'}_{\alpha \beta,\; \alpha' \beta'} \delta P^{\sigma'}_{\alpha' \beta'} (t).
\label{amat}
\end{equation}
Here $A^{\sigma \sigma'}_{\alpha \beta,\; \alpha' \beta'}$ is the spin-dependent stability matrix.
Each element of this stability matrix can be readily obtained from (\ref{tdhf-lin}). In general case it is sparse, complex and non-Hermitian matrix.
Now we need to find eigenvalues $\lambda_i$ of the stability matrix $A^{\sigma \sigma'}_{\alpha \beta,\; \alpha' \beta'}$ and analyze their real parts.
If   $\text{Re}(\lambda_i) <0$ for all eigenvalues in the spectrum, then the fixed point $\overline{P}^\sigma_{\alpha \beta}$ is asymptotically stable as $t\rightarrow \infty$ and, therefore, it is the true  steady state of the system. If at least one eigenvalue has a positive real part, the solution becomes  unstable along this mode and it is not a steady state.\cite{mikhailov,Strogatz}
The purely imaginary  eigenvalues  of stability matrix correspond to  the periodically oscillating  fixed points
which may be relevant to dynamical picture of  Coulomb blockade regime.\cite{PhysRevLett.104.236801}
If one of $\lambda_i$ becomes zero, the system undergoes zero-eigenvalue bifurcation which can be saddle-node, transcritical, or pitchfork type bifurcation.\cite{Strogatz}

\section{Example calculations}

\begin{figure}[t!]
 \begin{centering}
\includegraphics[width=0.8\columnwidth]{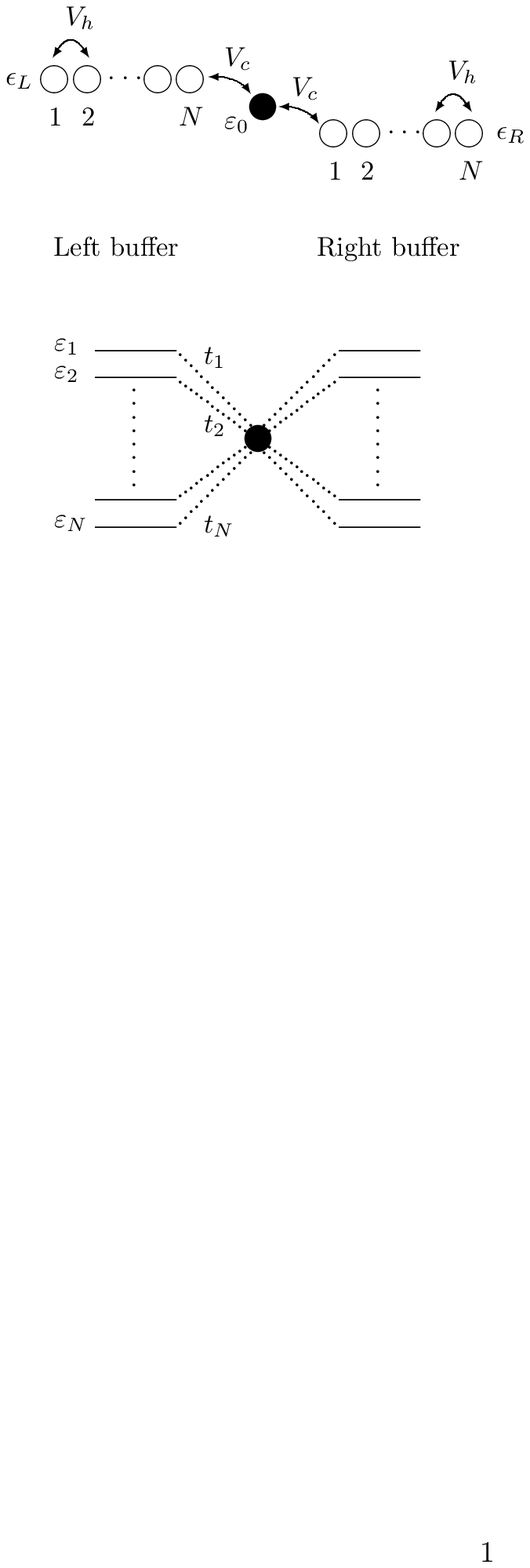}
 \end{centering}
 \caption{Schematic illustration of the model system used in the electron transport calculations. In the upper part of the figure only the first left $\epsilon_1~(1\in L)$
 and the last right $\epsilon_N~(N\in R)$ atoms  from the buffer zones are attached to the environment. After the diagonalization of the buffer zone Hamiltonian, each energy
 level $\varepsilon_b$  is connected to the dissipators and the molecule  (lower part of the figure). }
 \label{fig1}
\end{figure}

To illustrate the theory we  consider electron transport through a molecule with
a spin-degenerate single level with local Coulomb interaction
(so called Anderson model). The Hamiltonian is given by
 \begin{equation}
 \label{And_embedded}
   H = H_M   - \sum_{\sigma b} t_b (a^\dag_\sigma a_{b\sigma } + h.c)+
   \sum_{\sigma b} \varepsilon_b a^\dag_{b \sigma }a_{b \sigma },
   \end{equation}
  where the molecular Hamiltonian is
  \begin{equation}
   H_M = \varepsilon_0 \sum_\sigma a^\dag_\sigma a_\sigma + U a^\dag_\uparrow a_\uparrow a^\dag_\downarrow a_\downarrow.
\end{equation}

In our calculations left and right buffer zones are  modeled as a finite chain of $N$  atoms, characterized by the hopping parameter $V_h$ and the on-site energy $\epsilon_{L,R}$
(Fig.~\ref{fig1}). Thus, the energy spectrum of the each buffer is given by
 \begin{align}
  & \varepsilon_b = \epsilon_{L,R} + 2V_h  \cos\left(\frac{\pi b}{N+1}  \right),~~~b=1,\ldots,N.
 \end{align}
If  $V_c$ is the spin independent coupling between the molecule and the edge buffer site, then  the tunneling matrix elements in  Eq.~\eqref{And_embedded} are
  \begin{align}
  & t_b =  V_c \sqrt{\frac{2}{N+1}}\sin\left(\frac{\pi b}{N+1}  \right),~~~b = 1,\ldots, N.
 \end{align}
The parameter  $\gamma_b$  in Lindblad master equation is taken to be equal to the distance between neighbor energy levels in the buffer zones,
i.e.,  $\gamma_b = \varepsilon_b -  \varepsilon_{b+1}$.

In our calculations we use the same parameters as in~\cite{leeuwen10}. Namely,  $V_h=-0.5$,  $V_c=-0.35$, and the molecular orbital energy is $\varepsilon_0 = 0 $.  The buffer
on-site energies are shifted by the applied voltage bias ($V=V_L -V_R$): $\epsilon_{L,R} = 0.3 + V_{L,R}$, where $V_L=1.5$ and $V_R=0.0$.
The leads are half-filled so that the Fermi levels of lead $L,R$ are positioned at $\epsilon_{L,R}$. The inverse temperature is $\beta = 90$.
We use $N=400$ and this choice will be justified below.
Here we keep $U$ as a variable quantity and illustrate how the number and properties of nonequilibrium fixed points depend on the strength of electron-electron interaction.

\begin{figure}[t!]
 \begin{centering}
\includegraphics[width=0.9\columnwidth]{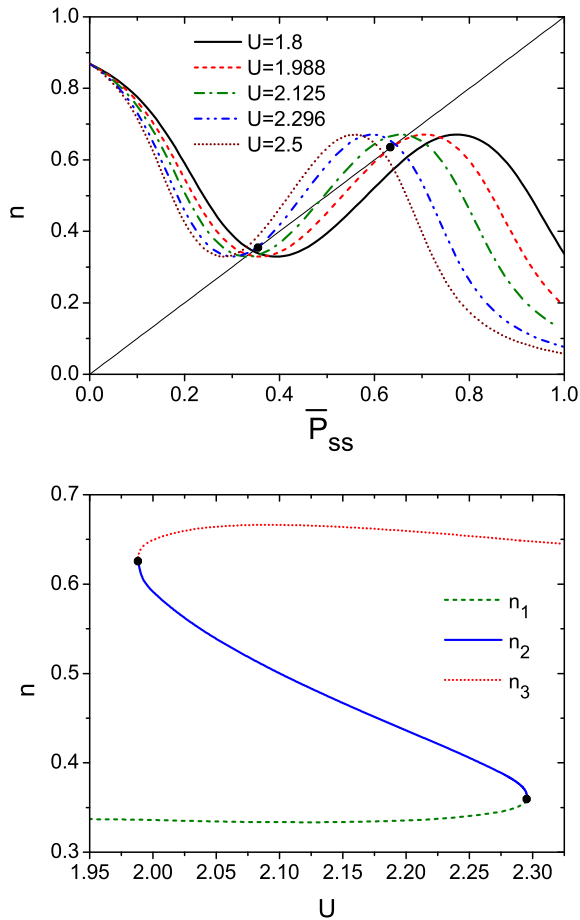}
\caption{Upper panel: The graphical solution of stationary nonequilibrium Hartree-Fock equations (\ref{hf-anderson}) for different valued of $U$. Lower panel: the fixed point electron densities
$n_k = \overline{P}^{(k)}_{ss}$ ($k=1,2,3$) as  functions of $U$. Black circles indicates
saddle-node bifurcation points.}
\label{fig2}
 \end{centering}
\end{figure}

The time-dependent Hartree-Fock  dynamics for the model Hamiltonian (\ref{And_embedded})  is fully characterized by the following (spin independent) single-particle density matrix:
\begin{align}
&P_{ss}(t) =  \langle a^\dag_{ \sigma} a_{\sigma} \rangle_t ,
\notag\\
&P_{sb}(t) =  \langle  a^\dag_{ b \sigma} a_{\sigma}\rangle_t , \; P_{bb'}(t) =  \langle a^\dag_{ b' \sigma} a_{b \sigma}\rangle_t.
\end{align}
To find the fixed point densities, $\overline{P}_{\alpha\beta}$, we solve the nonequilibrium stationary Hartree-Fock equations (i.e. the system of equations (\ref{tdhf}) with the time-derivatives of the density matrix set to zero):
\begin{align}
& \sum_{b} t_{b}\left (\overline{P}_{sb} - \overline{P}_{bs} \right)  =0,
     \notag\\
& t_b  \overline{P}_{ss}+ (\varepsilon_0 +  U \overline{P}_{ss} - E^*_b)\overline{P}_{sb} - \sum_{b'}t_{b'}\overline{P}_{b'b} =0,
     \notag\\
& t_b\overline{P}_{sb'} - t_{b'}\overline{P}_{bs} - (E_b - E^*_{b'})\overline{P}_{bb'} = 2i\delta_{bb'} f_b\gamma_b.
\label{hf-anderson}
   \end{align}
These Hartree-Fock equations are nonlinear with respect to the fixed point electron density in the molecule, $n = \overline{P}_{ss}$.

By numerical solution of the Hartree-Fock equations  we have found that
there is a range of the Coulomb interaction strength parameters,   $U$,  for which Eq.~\eqref{hf-anderson} have multipole fixed point solutions
$\overline{P}^{(k)}_{\alpha\beta}$ ($k=1,2,\ldots$).
In the upper panel of Fig.~\ref{fig2}  we show the graphical solution of Eq.~(\ref{hf-anderson}) for the different values of $U$.
In this plot the crossings of the straight and curved lines give molecular  densities $n = \overline{P}_{ss}$ corresponding to different fixed point solutions of Eq.~(\ref{hf-anderson}).
We see that three cases are possible depending on the value of $U$: when $1.988 < U < 2.296$ there exist three fixed points; at the ends of the
interval we have two fixed points; and outside the interval there is only one fixed point.
Hereafter we will number fixed point molecular  densities in ascending order, i.e., $n_1\le n_2\le n_3$.

In the lower panel of Fig.~\ref{fig2} we show how the molecular  densities $n_k$  depend on the Coulomb interaction strength.
The "middle" density  $n_2$ demonstrates the strong $U$ dependency and it exists only when $1.988 < U < 2.296$. The
fixed point corresponding to $n_2$ collides with that corresponding to $n_3$ ($n_1$) and annihilate it at the left (right) end of this interval for  $U$. This is so-called saddle-node bifurcation point~\cite{Strogatz} (indicated by black circles in Fig.~\ref{fig2}).

\begin{figure}[t!]
 \begin{centering}
\includegraphics[width=0.9\columnwidth]{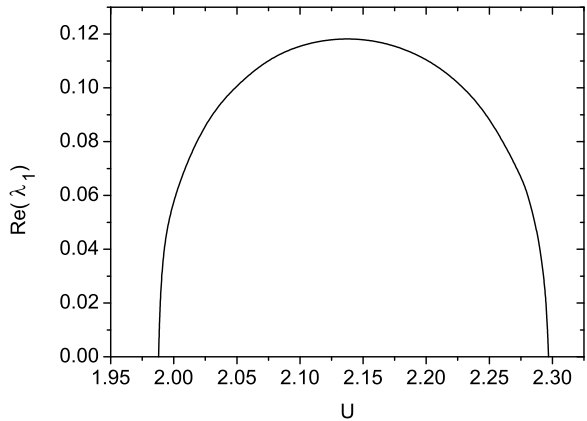}
\caption{The dependence of $\mathrm{Re}(\lambda_1)$ as a function of $U$ for the unstable nonequilibrium fixed point. }
\label{fig3}
 \end{centering}
\end{figure}

Let us understand which of  these three fixed point solutions are asymptotically stable (i.e.,  nonequilibrium steady states).
For this purpose  we  construct the stability matrix~\eqref{amat} for our model Hamiltonian.
The only non-zero matrix elements of the stability matrix are
\begin{eqnarray}
 && A_{ss,sb}   = -it_b,\;  A_{ss,bs} = i t_b, \; A_{sb,b'b} = i t_{b'}
  \nonumber\\
 && A_{sb,ss}  = -i(t_b + U \overline{P}_{sb}), \; A_{sb,sb}=-i[(\varepsilon_0 + U \overline{P}_{ss})  - E^*_b],
  \nonumber\\
 && A_{bs,ss}  = i(t_b + U \overline{P}_{bs}), \; A_{bs,bs}=i[(\varepsilon_0 + U\overline{P}_{ss}) - E_b],
  \nonumber\\
 && A_{bs,bb'} =  -it_{b'},\;  A_{bb',sb'} = it_b,  A_{bb',bs} = -it_{b'},\;
 \nonumber
 \\
  && A_{bb',bb'} = -i(E_b - E^*_{b'}).
\end{eqnarray}
The dimension of the stability matrix is $(1+2N)^2$ and it is sparse.  For a given fixed point   $\overline{P}^{(k)}_{\alpha\beta}$  we compute numerically   first few eigenvalues of the corresponding stability matrix  with  the largest real parts ($\mathrm{Re}(\lambda_1)\geq \mathrm{Re}(\lambda_2)\geq\ldots$).
For this aim we used the  ARPACK nonsymmeteric sparse eigenvalue solver.\cite{arpack}
We find  that the  stability matrices resulting from $\overline{P}^{(1)}_{\alpha\beta}$ and  $\overline{P}^{(3)}_{\alpha\beta}$
do not have eigenvalues with positive real parts, therefore the corresponding  fixed points are stable.
Contrary,  the stability matrix resulting from $\overline{P}^{(2)}_{\alpha\beta}$ always  has at least one eigenvalue $\lambda_1$ with a positive real part, i.e., the corresponding  fixed point is unstable.

In Fig.~\ref{fig3} we plot $\mathrm{Re}(\lambda_1)$ as a function of $U$ for the unstable fixed point $\overline{P}^{(2)}_{\alpha\beta}$. The obtained curve is symmetric with respect to $U = 2.143$ where $\mathrm{Re}(\lambda_1)$ reaches its maximum value. At $U=1.988$ and $U=2.296$, when the saddle-node bifurcation points are reached, eigenvalue $\lambda_1$ becomes zero.
Thus, our stability analysis is consistent with  the dynamical observations of~\cite{leeuwen10}  that the "middle" fixed point can not be reached by time-dependent propagation of the NEGF Kadanoff-Baum equation. We do not observe  purely imaginary eigenvalues of the stability matrix, therefore we rule out the existence of periodically oscillating fixed points in non-equilibrium Hartree-Fock approximation for the Anderson model.

In Table~\ref{table1} we compare
the calculated ($N=200,~400,~600$) fixed point densities of the molecule with the exact Hartree-Fock densities obtained by the NEGF method. The latter are the solutions of the
following  equation:\cite{haug-jauho}
\begin{equation}
\label{n}
  n= \frac{1}{\pi} \int d\omega \frac{\Gamma_L(\omega) f_L(\omega)+ \Gamma_R(\omega) f_R(\omega)}{(\omega-\varepsilon - U n  -\Lambda (\omega))^2 +(\Gamma(\omega))^2}.
\end{equation}
Here  $f_{L,R}(\omega) = [1 + e^{(\omega -\mu_{L,R})/T}]^{-1}$ is the Fermi-Dirac electron distribution in the left and right electrodes, and
$\Lambda=\Lambda_{L}+\Lambda_{R}$, $\Gamma=\Gamma_L+\Gamma_R$ are the real and imaginary parts of the leads self-energy.  For the tight binding
electrodes, that we consider, the self-energy is given by
\begin{align}
\label{SelfEnergy}
&\Sigma_{L,R} (\omega)= \Lambda_{L,R}(\omega)  - i\Gamma_{L,R}(\omega)
\notag
\\
&=\frac{V^2_h}{2V_c^2}\left\{
\begin{array}{c}
  \omega_{L,R} -   \sqrt{\omega^2_{L,R} - 4V^2_c},~~\omega_{L,R} >   2|V_c|  \\
  \omega_{L,R} +   \sqrt{\omega^2_{L,R} - 4V^2_c},~~\omega_{L,R} <-  2|V_c|\\
  \omega_{L,R} - i \sqrt{4V^2_c - \omega^2_{L,R}},~~|\omega_{L,R}| < 2|V_c|
\end{array}\right.
 \end{align}
where $\omega_{L,R} = \omega - \epsilon_{L,R}$. As seen from the table the larger $N$ the better  our method  reproduce the exact result.
For $N=400$ our method quite well reproduce the exact result (the maximum deviation from the exact results is under 2\%),
therefore our choice of the  size  of the buffer zone is justified  and makes our Lindblad-type kinetic equation numerically exact for  electron transport calculations.

\begin{widetext}
\begin{center}
\begin{table}[t!]\label{table1}
\caption{ Self-consistent electron densities on the molecule $n_k = \overline{P}_{ss}^{(k)}$, ($k=1,2,3)$  for nonequilibrium fixed points computed by the Lindblad kinetic equation with  different size of the buffer zones ($N=200,400, 600$)
and by the NEGF method.}
\begin{tabular}{|c||c|c|c||c|c|c||c|c|c||c|c|c|}
\hline
 U & \multicolumn{3}{|c||}{$N = 200$} & \multicolumn{3}{|c||}{$N = 400$} & \multicolumn{3}{|c||}{$N = 600$}  &  \multicolumn{3}{|c|}{NEGF}
\\ \hhline{|~|---||---||---||---|}
     & $n_1$ & $n_2$ & $n_3$ & $n_1$ & $n_2$ & $n_3$ & $n_1$ & $n_2$ & $n_3$ & $n_1$ & $n_2$ & $n_3$
\\ \hhline{|-||---||---||---||---|}
2.0  & 0.34 & 0.60 & 0.64 & 0.34 & 0.59 & 0.65 & 0.34 & 0.59 & 0.65 & 0.33  & 0.58  & 0.66  \\
2.05 & 0.34 & 0.54 & 0.66 & 0.33 & 0.54 & 0.67 & 0.33 & 0.54 & 0.67 & 0.33  & 0.54  & 0.67 \\
2.10 & 0.34 & 0.50	 & 0.66 & 0.33 & 0.50 & 0.67 & 0.33 & 0.50 & 0.67 & 0.33 & 0.50 & 0.67 \\
2.15 & 0.34 & 0.47 & 0.66 & 0.33 & 0.47 & 0.66 & 0.33 & 0.47 & 0.67 & 0.33 & 0.47 & 0.67 \\
2.20 & 0.34 & 0.43 & 0.66 & 0.34 & 0.44 & 0.66 & 0.33 & 0.44 & 0.66 & 0.33 & 0.44 & 0.66\\
2.25 & 0.35 & 0.40 & 0.65 & 0.34 & 0.41 & 0.65 & 0.34 & 0.41 & 0.66 & 0.33 & 0.41 & 0.66\\
\hline
\end{tabular}
\end{table}
\end{center}
 \end{widetext}

\section{Conclusions}

We have presented a theoretical method to perform a stability analysis of multiple nonequilibrium fixed points, which appear in  self-consistent electronic transport calculations. We employed the Lindblad kinetic equation and  underlying approximations of this kinetic equation were alleviated by the use of  explicit buffer zone between the molecule and electronic electrodes. Nonequilibrium fixed point were obtained from the self-consistent solution of the
 kinetic equation in stationary limit. The asymptotic stability of these fixed  points were studied by linearizing of the nonlinear  kinetic equation. We obtained the non-Hermitian stability matrix from the linearized kinetic  equation and analyzed its spectrum.
If real parts of all eigenvalues of the stability matrix are negative, then the fixed point is asymptotically stable as $t\rightarrow \infty$ and it can be regarded as  steady state. If at least one eigenvalue has positive real part, the solution becomes  unstable and can not be the steady state.
We  obtained the explicit form of the stability matrix in Hartree-Fock and adiabatic time-dependent DFT approximations.
The method was applied to out of equilibrium Anderson model which yields three nonequilibrium fixed points under certain choice of parameters  in nonequilibrium Hartree-Fock approximation. We performed the stability analyse of these fixed points and demonstrated that one fixed point is asymptotically unstable whereas the other fixed points correspond to  physical steady states.

\begin{acknowledgments}
This work has been supported by the Francqui Foundation, Belgian Federal Government under the Inter-university Attraction Pole project NOSY  and
 Programme d'Actions de Recherche Concert\'ee de la Communaut\'e francaise (Belgium) under project "Theoretical and experimental approaches to surface reactions".
\end{acknowledgments}

\appendix

\section{Lindblad kinetic equation for embedded molecule}

Let  us begin with the Liouville equation for the total density matrix $\chi(t)$
\begin{equation}
i \dot{{\chi}}(t) = [{\cal H}, \chi(t)].
\end{equation}
We partition the Hamiltonian (\ref{h}) into two parts:
\begin{equation}
h =  H_M+  H_E + {H}_{B} + H_{SB}
\end{equation}
\begin{equation}
v= H_{EB}
\end{equation}

In the interaction representation the Liouville equation becomes
\begin{align}\label{kin2}
\dot\chi_I(t) &= \frac{1}{i}   [ v_I(t), \chi_I(0)]
\notag\\
& - \int_0^{t} d\tau [[ v_I(t), [ v_I(t-\tau),\chi_I(t-\tau)]],
\end{align}
where $\chi_I(t)= e^{ih t} \chi(t) e^{-iht}  $ and $v_I(t) = e^{ih t} v e^{-iht}$.
Now we introduce the density matrix for  embedded system (molecule plus buffer) by tracing out the environment degrees of freedom
\begin{equation}
\rho(t) = \text{Tr}_E  \chi(t)
\end{equation}
We assume that the total density matrix can be factorized and  environmental degrees of freedom propagate in time as there were no interaction  with the buffer
(Born approximation):
\begin{equation}
\chi_I(t) = \rho_I(t) \rho_E.
\end{equation}
The  density matrix for the environment is taken in the grand canonical ensemble form
\begin{equation}
\rho_E \sim e^{- \sum\limits_{\sigma, k\in L} \beta_{L} (\varepsilon_{k} - \mu_{L}) a^{\dagger}_{k\sigma} a_{k\sigma}} e^{- \sum\limits_{\sigma, k\in R}  \beta_{R} (\varepsilon_{k} - \mu_{R}) a^{\dagger}_{k\sigma}  a_{k\sigma}}.
\end{equation}
Setting $ \mu_{L} \ne \mu_{R} $ and/or $ \beta_{L} \ne \beta_{R} $  in the environment, we drive the system out of equilibrium.

After tracing, Eq.~(\ref{kin2}) becomes
\begin{equation}
\dot \rho_I(t) = -\int_0^t d\tau    \text{Tr}_E \left[ [ v_I(t), [ v_I(t-\tau),\rho_I(t-\tau) \rho_E] \right].
\end{equation}
Here, we have assumed that $\text{Tr}_E[v_I(t), \chi_I(0)]=0$.

Using the explicit expression for the Hamiltonian $h$ one can easily demonstrate that
\begin{align}
e^{ih t} a_{k \sigma} e^{-iht} &=  e^{-i \varepsilon_{k } t} a_{k \sigma},
\notag\\
e^{ih t} a_{b \sigma} e^{-iht} &= e^{-i \varepsilon_b t} a_{b \sigma } + O(1/N_B).
\end{align}
where $N_B$ is the number of discrete single particle levels (i.e. the size) of the buffer zone.
Therefore by choosing the large enough buffer zone, we may assume that  buffer single-particle states evolve in time as free states.
As a result, $v_I(t)$ takes the form
\begin{equation}
  v_I(t) = \sum_{\sigma bk} (v_{bk}(t)a^\dag_{b\sigma}a_{k\sigma} + h.c.).
\end{equation}
where $v_{bk}(t)= v_{bk} e^{i(\varepsilon_k - \varepsilon_b)t}$.

Now, the kinetic equation for the embedded system density matrix becomes
\begin{widetext}
\begin{align}\label{kin}
  \dot \rho_I(t) = -\int_0^t d\tau \sum_{\sigma k b b'} \Bigl\{v_{bk}(t)v^*_{b'k}(t-\tau)
  \Bigr[(1-f_k)a^\dag_{b\sigma}[a_{b'\sigma},\rho_I(t-\tau)] - f_k [a_{b'\sigma},\rho_I(t-\tau)]a^\dag_{b\sigma}\Bigl] +
  \notag\\
  v^*_{bk}(t)v_{b'k}(t-\tau)
  \Bigr[f_k a_{b\sigma}[a^\dag_{b'\sigma},\rho_I(t-\tau)] - (1-f_k) [a^\dag_{b'\sigma},\rho_I(t-\tau)]a_{b\sigma}\Bigl]\Bigr\} .
\end{align}
\end{widetext}
Here $ f_{k\in L/R} = \text{Tr}_E (\rho_E a^{\dag}_{k\sigma }  a_{k\sigma}) =[1+ e^{\beta_{L/R}(\varepsilon_{k} - \mu_{L/R}) }]^{-1}$
Assuming that the environment relaxation time is very fast we can extend the integration range to $+\infty$ and $\rho(t-\tau) \simeq \rho(t)$ (Markov approximation).
Finally, in the rotating wave approximation rapidly oscillating terms proportional to $\mathrm{\exp}[i(\varepsilon_b - \varepsilon_b')t]$ for $\varepsilon_b \ne \varepsilon_b'$
are neglected. Then,   the kinetic equation (\ref{kin}) becomes the standard Lindblad type master equation (\ref{lindblad}).

The obtained Lindblad master equation describes the time evolution of the open embedded system preserving the probability and the positivity of the density
matrix. Open boundary conditions are introduced via non-Hermitian dissipative part of Eq.\eqref{lindblad}, $\hat \Pi \rho(t)$, which represents the
influence of environment on the system.  The applied bias potential enters into Eq.\eqref{lindblad}
via fermionic occupation numbers $f_b~(b\in L,R)$ which depend on the chemical potential in the light
and right electrodes.

\section{Stability matrix for nonequilibrium self-consistent DFT electron transport calculations}

Leaving apart the conceptual questions about the use of ground state DFT for self-consistent electronic transport calculations, we can say that
the only difference in practical computations between nonequilibrium time-dependent Hartree-Fock discussion (Section II) and adiabatic time-dependent Kohn-Sham DFT is that
the time-dependent Fock matrix (\ref{fock-matrix})  becomes
\begin{equation}
F_{nm}(t)=  T_{nm} + \int d{\bf r} \phi_n ({\bf r})    v_{KS}^{\sigma}({\bf r},t) \phi_m ({\bf r}).
\end{equation}
Here
\begin{equation}
v_{KS}^{\sigma}  ({\bf r}, t) =  \int d{\bf r'} \frac{\rho({\bf r'},t)}{| {\bf r} - {\bf r'}|} + v_{xc}^\sigma [\rho]({\bf r},t),
\end{equation}
where $v_{xc}^\sigma [\rho]({\bf r},t)$ is the exchange-correlation potential.
Then the system of equations (\ref{tdhf}) and (\ref{tdhf-lin}) remains the same,
but the steady state Fock matrix (\ref{fst}) becomes
\begin{equation}
\overline{F}_{nm}=  T_{nm} + \int d{\bf r} \phi_n ({\bf r})  v_{KS}^{\sigma}[\overline{\rho}] ({\bf r}) \phi_m ({\bf r})
\end{equation}
and variation (\ref{deltaf}) is
\begin{equation}
\delta {F}_{nm}(t)=  \int d{\bf r} \phi_n ({\bf r})   \delta  v_{KS}^{\sigma}({\bf r},t) \phi_m ({\bf r}),
\end{equation}
where $\delta  v_{KS}^{\sigma}$ is given by the standard expression of linear response  time-dependent DFT
\begin{align}
&\delta v^{\sigma}_{KS} ({\bf r}, t) = \int d{\bf r'} \frac{\delta \rho({\bf r'},t)}{| {\bf r} - {\bf r'}|}
\notag\\
& + \sum_{\sigma'} \int d{\bf r'} f_{xc}(\sigma {\bf r}, \sigma' {\bf r'} ) \delta \rho_{\sigma'}({\bf r'},t).
\end{align}
The exchange-correlation response kernel
$f_{xc}(\sigma {\bf r}, \sigma' {\bf r'} ) $
is given in the usual adiabatic approximation,\cite{gross90,casida} i.e., the
exchange-correlation contribution is taken to be simply the
second derivative of the static ground state exchange-correlation
energy $E_{xc}$ with respect to the fixed point spin density $\overline{\rho}_\sigma({ \bf r})$:
\begin{equation}
f_{xc}(\sigma {\bf r}, \sigma' {\bf r'} ) = \delta({\bf r} - {\bf r'}) \frac{\delta E_{xc}[\overline{\rho}]}{\delta \overline{\rho}_{\sigma'}({\bf r'}) \delta \overline{\rho}_{\sigma}({\bf r})}
\end{equation}
Therefore, the stability analysis of nonequilibrium fixed points for self-consistent DFT electronic transport calculations can be readily performed  within standard adiabatic time-dependent density functional response theory.
\cite{doltsinis2000563,furche:7433,doltsinis:144101}



\end{document}